%% file: main.tex
\documentclass[10pt,journal]{IEEEtran}
\usepackage[letterpaper,top=1in,bottom=1in,left=0.7in,right=0.7in,columnsep=0.2in]{geometry}
\usepackage{amsmath,amssymb,mathtools}
\usepackage{graphicx}
\usepackage{booktabs}
\usepackage{placeins}
\usepackage{dblfloatfix}
\usepackage{url}
\usepackage[hidelinks]{hyperref}
\hypersetup{pdftitle={Safe Two-Component Gaussian Mixture Reduction for Linear Error Propagation},
  pdfauthor={Penggao Yan, Baoshan Song, Yuan Li, Li-Ta Hsu}}
\graphicspath{{figures/}}

\title{Safe Two-Component Gaussian Mixture Reduction for Linear Error Propagation}
\author{Penggao Yan, Baoshan Song, Yuan Li, and Li-Ta Hsu%
\thanks{All authors are with The Hong Kong Polytechnic University. Corresponding author: Penggao Yan (penggao.yan@polyu.edu.hk).}}

\begin{document}
\maketitle

\begin{abstract}
\input{sections/00-abstract}
\end{abstract}


\input{sections/01-introduction}
\input{sections/02-problem}
\input{sections/03-construction}
\input{sections/04-sequential}
\FloatBarrier
\setlength{\dbltextfloatsep}{0pt}
\setlength{\dblfloatsep}{0pt}
\input{sections/05-results}
\setlength{\dbltextfloatsep}{12pt}
\setlength{\dblfloatsep}{12pt}
\restoregeometry
\input{sections/06-conclusion}
\FloatBarrier

\section*{Acknowledgment}
OpenAI Codex running through ManuForge was used to organize the approved source materials and to draft and typeset Sections~I--VI, the appendices, tables, figures, and LaTeX source. The system did not generate new experimental evidence; source verification and scientific decisions remain tied to the cited records.

\appendices
\input{sections/07-appendices}
\FloatBarrier

\bibliographystyle{IEEEtran}
\bibliography{references}

\end{document}

%% file: sections/00-abstract.tex
Protection-level (PL) computation must remain conservative when measurement errors are heavier-tailed than a nominal Gaussian. A finite zero-mean Gaussian scale mixture can represent a concentrated core and a broader tail more flexibly, but explicit propagation retains all mode combinations and rapidly increases the cost of posterior and quantile calculations. This paper develops a calculable two-component reduction for centered finite Gaussian mixtures. The construction fixes the largest input scale, solves one scalar tail-contact equation for the internal scale, and recovers the mixture weight algebraically. A principal-representation result gives the all-threshold tail overbound, while the contact construction is exact at the selected protection level. We then apply the reduction after each independent scalar convolution. The resulting induction preserves the all-threshold tail order; a separate convolution identity explains how a nontrivial early tail gap becomes a strict terminal protection-level gap. In synthetic evaluations, the method keeps at most two propagated components. Across all 16 supplementary cases, its protection levels are lower than those from a fixed-budget comparator that performs a two-component groupwise upward-variance merge. In the 48-case main matrix, the $N=12$ sequential-C wall time is approximately $8.67$ ms, compared with $167.47$ ms for exact propagation in the recorded environment. The results quantify the compact error-representation trade-off in the stated propagation setting.

%% file: sections/01-introduction.tex
\section{Introduction}
\label{sec:introduction}

In Global Navigation Satellite System (GNSS) integrity monitoring, a scalar projection of a linearized navigation solution uses a protection level as the threshold whose two-sided exceedance probability equals the allocated risk level. The model used for that calculation therefore has to provide a conservative upper bound on tail probability, not only a central variance. This requirement becomes difficult when measurement errors are heavier-tailed than a nominal Gaussian~\cite{braff2005}. Gaussian overbounds often enlarge their scale to accommodate observed tail probabilities, which places the concentrated part of the error mass and the broader tail under one parameter and can inflate the resulting protection level~\cite{larson2019,yan2024pgo}. Finite Gaussian mixtures provide a more flexible description by assigning different scales and weights to these two features; they have consequently entered position-error and protection-level calculations in integrity monitoring and related navigation analyses \cite{gao2022gmm,perea2019multigaussian,hou2025carrier}.

The present study takes such a centered Gaussian mixture as a given input model and addresses its reduction during linear propagation. This is distinct from the modeling step in the principal Gaussian overbound work, which constructs a Gaussian error bound for a heavy-tailed distribution \cite{yan2024pgo}.

The flexibility of a mixture creates a second problem when the model is propagated. If each independent measurement retains several modes, the joint representation contains one component for every mode combination. In the Bayesian receiver autonomous integrity monitoring (RAIM) calculation of Blanch, Walter, and Enge, each conditional mode combination carries its own covariance, position estimate, and posterior weight, so the number of combinations directly controls the number of posterior calculations \cite{blanch2008position}. In the mixture-propagation route of Gao \emph{et al.}, the same combinations are convolved into a position-domain mixture and used to evaluate tail probabilities and quantiles \cite{gao2022gmm}. These are different downstream calculations, but they expose the same scaling pressure: a mixture that is useful as an error model becomes difficult to retain as an explicit propagation state. Related GBAS and PPP applications keep mixture components in solution-set grouping or component-specific filtering, with the former using risk allocation and the latter deriving protection levels from the combined state estimate and covariance~\cite{wang2023gbasgmmpl,wang2022pppgmmpl}. Compact discrete envelopes and Gaussian sample bounds likewise address the computational burden of non-Gaussian convolutions~\cite{rife2012navden,blanch2019}.

The reduction is formulated to preserve the error event used by integrity monitoring, while the protection-level increase introduced by the reduction remains measurable at the selected risk level. We write this requirement using the CDF overbounding semantics used in GNSS integrity analysis: the candidate must not assign a smaller probability to either side of a large-error threshold than the input model \cite{rife2006paired,blanch2019}. For the centered symmetric class considered here, this is equivalently an all-threshold order on the two-sided absolute tail. The protection level at a fixed $\alpha$ then supplies the application-facing measure of the remaining conservatism. Classical mixture-reduction objectives such as Kullback--Leibler (KL) divergence address distributional approximation~\cite{runnalls2007klgmr}. Other reduction methods instead use integral squared error and its normalized form~\cite{crouse2011gmr}, clustering~\cite{schieferdecker2009gmrc}, or two-component moment matching with normalized integral squared error~\cite{yokoyama2016twogmm}. Composite Transportation Divergence (CTD) instead optimizes a transport-based discrepancy between mixtures, while upward variance merging imposes a conservative direction; these objectives do not by themselves supply the fixed-level, all-threshold construction needed when a mixture is retained as an integrity error state~\cite{zhang2024ctd,gao2022gmm}. However, for an integrity error state we seek a construction that preserves the tail bound at every threshold and, at the selected risk level, matches the input protection level; we apply principal-representation tools to this specific overbounding problem.

For a centered finite Gaussian mixture, we keep the largest input scale as the wide component, determine an internal scale from the current protection level and tail slope, and recover the wide-component weight from the contact equations. The resulting two-component representation is all-threshold safe and has the same protection level as its current input in exact arithmetic. We then use it as a state after each nonzero scalar convolution. Symmetric-unimodal convolution preservation gives the sequential safety induction, while a separate tail-gap identity shows why a locally exact reduction can still produce a strict final protection-level increase after a later nondegenerate convolution. The evaluation compares exact propagation, sequential C reduction, sequential peel, terminal C, and a same-budget upward-merge comparator on the stated synthetic cases.

The contributions are threefold. First, we give the Gaussian-tail specialization of the classical principal-representation construction, including its fixed-largest-scale parameter equations, nondegeneracy conditions, and computable solution. Second, we establish sequential safety and identify a sufficient condition under which an early nontrivial reduction produces a strict terminal protection-level gap. Third, we quantify the protection-level, intermediate-size, and runtime trade-offs on 48 main cases, 12 held-out cases, eight boundary cases, and a separate 16-case same-budget supplement. The two-component budget retains separate Gaussian scales and weights in a compact propagated state. Sections~II--IV define the model, construct the reduction, and analyze sequential propagation. Section~V reports the numerical evidence, and Section~VI concludes with the supported implications and two extensions.

%% file: sections/02-problem.tex
\section{Problem Formulation and Preliminaries}
\label{sec:problem}

\subsection{Centered finite Gaussian-mixture inputs}

We consider a scalar component of a linearized error-propagation model~\cite{nikiforov2019risk}. The $i$th input
error $X_i$ has a centered finite Gaussian-mixture density
\begin{equation}
 \begin{aligned}
 f_i(x)&=\sum_{k=1}^{K_i}w_{ik}\,\phi(x;0,\sigma_{ik}^{2}),\\
 &\qquad w_{ik}>0,\quad \sum_{k=1}^{K_i}w_{ik}=1,\quad \sigma_{ik}>0,
 \end{aligned}
 \label{eq:input-gmm}
\end{equation}
and $X_1,\ldots,X_N$ are mutually independent. Here $\phi(\cdot;0,\sigma^2)$ denotes the
Gaussian density with zero mean and variance $\sigma^2$. The centering and positive-scale
assumptions make every input density symmetric and strictly decreasing on the positive
half-line. They also give a common tail representation for the construction in Section~III.
Equivalently, this input chooses a positive Gaussian scale from a finite set, giving a
discrete normal scale mixture~\cite{andrewsmallows1974scale,west1987scalemixtures}.

The propagated scalar error is
\begin{equation}
 E=\sum_{i=1}^{N}s_iX_i,\qquad s_i\in\mathbb{R},
 \label{eq:linear-error}
\end{equation}
where at least one coefficient is nonzero. Let $H_i$ denote the probability distribution of
$s_iX_i$. For $s_i\neq0$,
\begin{equation}
 H_i=\sum_{k=1}^{K_i}w_{ik}\,\mathcal{N}\!\left(0,s_i^2\sigma_{ik}^2\right),
 \label{eq:scaled-input}
\end{equation}
where the sign of $s_i$ disappears because the input is symmetric. If $s_i=0$, $H_i$ is the
point mass at zero, denoted by $\delta_0$; this term is an identity under convolution and is
skipped by the recursive implementation.

Independence makes convolution the exact propagation operation, as in mechanistic non-Gaussian
GBAS propagation through a linear position solution~\cite{dautermann2012nongaus}. If $F_N$ is the distribution
of $E$, then
\begin{equation}
 F_N=H_1*H_2*\cdots*H_N.
 \label{eq:exact-convolution}
\end{equation}
For the nonzero-index set $I=\{i:s_i\neq0\}$, each mode tuple has weight
\begin{equation}
 w_{\boldsymbol{k}}=\prod_{i\in I}w_{ik_i},\qquad
 \sigma_{\boldsymbol{k}}^2=\sum_{i\in I}s_i^2\sigma_{ik_i}^2,
 \label{eq:mode-combination}
\end{equation}
and the nominal number of tuples is $\prod_{i\in I}K_i$. Equal variances can be merged, so
the number of distinct scales can be smaller, but the product is the relevant worst-case
growth of an explicit representation.

\subsection{Two-sided tails, protection levels, and safe propagation}

The following definitions fix the tail and protection-level quantities used by the reduction.

Let $Z\sim\mathcal{N}(0,\sigma^2)$. The two-sided tail for $x\geq0$ is given by the following equation.
\begin{equation}
 Q_\sigma(x)=\Pr(|Z|>x)
 =\operatorname{erfc}\!\left(\frac{x}{\sqrt{2}\,\sigma}\right).
 \label{eq:gaussian-tail}
\end{equation}
For a centered mixture $f$, the corresponding absolute-tail function is
\begin{equation}
 T_f(x)=\Pr_f(|X|>x)=\sum_{k=1}^{K}w_kQ_{\sigma_k}(x).
 \label{eq:mixture-tail}
\end{equation}
It satisfies $T_f(0)=1$, tends to zero as $x$ increases, and has derivative
\begin{equation}
 T_f'(x)=-\sqrt{\frac{2}{\pi}}\sum_{k=1}^{K}\frac{w_k}{\sigma_k}
 \exp\!\left(-\frac{x^2}{2\sigma_k^2}\right)<0,\qquad x>0.
 \label{eq:tail-derivative}
\end{equation}
Therefore, for a fixed two-sided tail-risk level $0<\alpha<1$, the protection level (PL) is the
unique positive solution
\begin{equation}
 \operatorname{PL}_\alpha(f)=x_\alpha
 \quad\Longleftrightarrow\quad T_f(x_\alpha)=\alpha.
 \label{eq:pl-definition}
\end{equation}
In this scalar setting, the interval $[-\operatorname{PL}_\alpha(f),
\operatorname{PL}_\alpha(f)]$ contains probability at least $1-\alpha$ under $f$.

To compare an exact mixture with a reduced representation, we introduce the notation
$f\preceq g$ for the all-threshold safety relation
\begin{equation}
 f\preceq g\quad\Longleftrightarrow\quad T_g(x)\geq T_f(x)\quad\text{for every }x\geq0.
 \label{eq:tail-order}
\end{equation}
Let $F_f$ and $F_g$ denote the cumulative distribution functions (CDFs) of $f$ and $g$. For continuous zero-centered symmetric
distributions,~\eqref{eq:tail-order} is equivalent to the usual CDF overbound directions,
\begin{equation}
 F_f(x)\leq F_g(x)\;(x\leq0),\qquad
 F_f(x)\geq F_g(x)\;(x\geq0),
 \label{eq:cdf-order}
\end{equation}
because the positive and negative tails are equal. Since both tails are strictly decreasing,
\begin{equation}
 f\preceq g\quad\Longrightarrow\quad
 \operatorname{PL}_\alpha(f)\leq\operatorname{PL}_\alpha(g),
 \qquad T_f\!\left(\operatorname{PL}_\alpha(g)\right)\leq\alpha.
 \label{eq:order-pl}
\end{equation}
Thus a safe reduction cannot lower the protection level relative to its input; the increase is
the conservatism that the construction and experiments must quantify.

The target problem is to replace an explicit mixture by at most two centered Gaussian
components while preserving~\eqref{eq:tail-order}. The representation budget is fixed because
it controls the state passed to the next convolution; the construction itself is developed in
Section~III, and its recursive use is analyzed in Section~IV.

The centered Gaussian mixtures in~\eqref{eq:input-gmm} are symmetric unimodal densities. A
classical peakedness result states that if $f\preceq g$ and an independent kernel $\eta$ is
symmetric unimodal, then
\begin{equation}
 f*\eta\preceq g*\eta.
 \label{eq:convolution-order}
\end{equation}
This is the bridge from a single safe reduction to a sequential propagation guarantee. The
result is attributed to the classical peakedness literature and the convolution theorem
summarized by Bickel--Lehmann and Shaked--Shanthikumar
\cite{bickel1976dispersion,shaked2007stochastic}.
The short integral proof specialized to the present mixture class is included in Appendix~\ref{app:derivations};
Section~IV uses~\eqref{eq:convolution-order} as a cited closure property. The construction
that enforces~\eqref{eq:tail-order} is introduced next; these definitions fix the model and
propagation relation used by the reduction.

%% file: sections/03-construction.tex
\section{Two-Component Overbound Construction}
\label{sec:construction}

\subsection{Candidate family and fixed-level contact}

After combining repeated scales and removing zero-weight terms, write the input as
\begin{equation}
 \begin{aligned}
 f(x)&=\sum_{k=1}^{K}w_k\phi(x;0,\sigma_k^2),\\
 &\qquad 0<\sigma_1<\cdots<\sigma_K,\quad w_k>0,\quad \sum_kw_k=1.
 \end{aligned}
 \label{eq:sorted-input}
\end{equation}
When $K\leq2$, the input already satisfies the representation budget. For $K\geq3$, we
consider
\begin{equation}
 \begin{aligned}
 g_{a,p}(x)&=(1-p)\phi(x;0,a^2)+p\phi(x;0,b^2),\\
 &\qquad 0<a\leq b,\quad 0\leq p\leq1,
 \end{aligned}
 \label{eq:c-candidate}
\end{equation}
and fix the wide scale to the input maximum, $b=\sigma_K$. This choice preserves the
largest available Gaussian scale and leaves the internal scale $a$ and the wide weight $p$ as
the two design degrees of freedom.

Here $a$ and $b$ are standard deviations in the same units as the propagated error, while
$p$ is dimensionless; the protection-level coordinate used below is expressed in those error units.

For a selected $\alpha$, the safe reduction problem is
\begin{equation}
 \min_{a,p}\ \operatorname{PL}_\alpha(g_{a,p})
 \quad\text{s.t.}\quad f\preceq g_{a,p}.
 \label{eq:c-optimization}
\end{equation}
The single Gaussian with scale $\sigma_K$ is feasible because $Q_{\sigma_k}(x)\leq
Q_{\sigma_K}(x)$ for every $k$ and $x\geq0$. Hence the feasible set is nonempty, and
~\eqref{eq:order-pl} gives $\operatorname{PL}_\alpha(f)$ as a lower bound for every safe
candidate. Let
\begin{equation}
 x_0=\operatorname{PL}_\alpha(f),\qquad T_f(x_0)=\alpha,\qquad
 D(x)=T_g(x)-T_f(x).
 \label{eq:contact-def}
\end{equation}
At a candidate attaining this lower bound, safety gives $D(x)\geq0$ and the contact point
is an interior minimum on the positive half-line. Consequently,
\begin{equation}
 D(x_0)=0,\qquad D'(x_0)=0.
 \label{eq:contact-equations}
\end{equation}
The two equations match the candidate and input tail value and tail slope at the selected
protection level; they do not by themselves establish all-threshold safety or strict curvature.
For positive weights on $K\geq3$ distinct scales, the resulting difference is positive for $y>0$,
$y\neq x_0$, and its second derivative at the contact has the strict sign recorded in
Appendix~\ref{app:derivations}.

\subsection{Scalar parameter equation}

Set $\lambda=1/(2\sigma^2)$, $A=1/(2b^2)=1/(2\sigma_K^2)$, and $c=1/(2a^2)$. Define
\begin{equation}
 U(\lambda)=\operatorname{erfc}(x_0\sqrt{\lambda}),\qquad
 V(\lambda)=-\frac{2\sqrt{\lambda}}{\sqrt{\pi}}
 \exp(-\lambda x_0^2).
 \label{eq:uv-functions}
\end{equation}
Here $U$ is the Gaussian tail at $x_0$ and $V$ is its derivative with respect to the
threshold $x$, evaluated at $x_0$. The contact equations are
\begin{equation}
 \begin{aligned}
 (1-p)U(c)+pU(A)&=\sum_kw_kU(\lambda_k),\\
 (1-p)V(c)+pV(A)&=\sum_kw_kV(\lambda_k),
 \end{aligned}
 \label{eq:uv-contact}
\end{equation}
where $\lambda_k=1/(2\sigma_k^2)$ and $\lambda_1>\cdots>\lambda_K=A$. Subtracting the
known endpoint terms and using $\lambda_K=A$ gives
\begin{equation}
 \begin{aligned}
 (1-p)h(c)&=\sum_{k<K}w_kh(\lambda_k),\\
 (1-p)\psi(c)&=\sum_{k<K}w_k\psi(\lambda_k),
 \end{aligned}
 \label{eq:hpsi-contact}
\end{equation}
with
\begin{equation}
 h(\lambda)=U(A)-U(\lambda),\qquad \psi(\lambda)=V(A)-V(\lambda).
 \label{eq:hpsi-def}
\end{equation}
Because $U$ is strictly decreasing and each light precision satisfies $\lambda_k>A$,
$h(\lambda_k)=U(A)-U(\lambda_k)>0$. For a candidate with $p\leq1$ and $c\geq A$,
the left factor $h(c)$ is also nonnegative; a nondegenerate contact therefore has
$1-p>0$ and $c>A$. Division by the first equation is legitimate. Define
\begin{equation}
 r(\lambda)=\frac{\psi(\lambda)}{h(\lambda)},\qquad
 R=\frac{\sum_{k<K}w_k\psi(\lambda_k)}{\sum_{k<K}w_kh(\lambda_k)}.
 \label{eq:r-R}
\end{equation}
The two unknowns in~\eqref{eq:uv-contact} have now been separated: $c$ is the unique root
of $r(c)=R$, and $p$ is recovered after that root is found.

\subsection{Existence, uniqueness, and weight recovery}

The endpoint derivatives are
\begin{equation}
 U'(\lambda)=-\frac{x_0}{\sqrt{\pi\lambda}}e^{-\lambda x_0^2}<0,\qquad h'(\lambda)=-U'(\lambda)>0.
 \label{eq:u-derivative}
\end{equation}
The Gaussian-tail identity gives
\begin{equation}
 \frac{\psi'(\lambda)}{h'(\lambda)}=\frac{V'(\lambda)}{U'(\lambda)}
 =\frac{1}{x_0}-2x_0\lambda,
 \label{eq:ratio-derivative}
\end{equation}
which is strictly decreasing in $\lambda$. With the common endpoint values
$h(A)=\psi(A)=0$ and $h'(\lambda)>0$, the monotone form of l'Hospital's rule therefore
shows that $r(\lambda)$ is continuous and strictly decreasing on $(A,\infty)$~\cite{pinelis2002lhospital}. Moreover,
\begin{equation}
 R=\frac{\sum_{k<K}w_kh(\lambda_k)r(\lambda_k)}{\sum_{k<K}w_kh(\lambda_k)}
 \label{eq:R-convex-combination}
\end{equation}
is a strict weighted average when at least two distinct light scales have positive weight.
Thus
\begin{equation}
 r(\lambda_1)<R<r(\lambda_{K-1}),\qquad
 c\in(\lambda_{K-1},\lambda_1),
 \label{eq:c-interval}
\end{equation}
and the intermediate-value theorem together with strict monotonicity gives one and only one
root. The recovered parameters are
\begin{equation}
 a=(2c)^{-1/2},\qquad b=\sigma_K,\qquad
 1-p=\frac{\sum_{k<K}w_kh(\lambda_k)}{h(c)}.
 \label{eq:parameter-recovery}
\end{equation}
Equivalently, the contact weight can be checked directly from
\begin{equation}
 p=\frac{\alpha-Q_a(x_0)}{Q_b(x_0)-Q_a(x_0)}.
 \label{eq:p-check}
\end{equation}
The strict concavity underlying the weight range follows from the auxiliary curve
$P(h(\lambda))=\psi(\lambda)$. Since
\begin{equation}
 P'(h(\lambda))=\frac{1}{x_0}-2x_0\lambda,\qquad
 P''(h(\lambda))=-\frac{2x_0}{h'(\lambda)}<0,
 \label{eq:P-concavity}
\end{equation}
strict Jensen comparison yields $w_K<p<1$ for a nondegenerate $K\geq3$ input. This
inequality is useful numerically because it separates the recovered wide weight from the input
wide weight and provides a direct range check.

\subsection{All-threshold safety}

The contact equations match the three weighted quantities represented by total mass, the tail
value at $x_0$, and the tail slope at $x_0$. They establish equality of value and slope at the
selected point, but not all-threshold safety or strict curvature. The Gaussian-tail
functions $\{1,-U(\lambda),V(\lambda)\}$ form the base Chebyshev system used by the
principal-representation result. For a comparison threshold $y>0$ distinct from $x_0$, adding
$-\operatorname{erfc}(y\sqrt{\lambda})$ gives the corresponding extended system on
$[A,\lambda_1]$. At $y=x_0$ the tail equality is already one of the base constraints, and at
$y=0$ it follows from normalization. Under the
positive-weight, distinct-scale, and $K\geq3$ conditions, the relevant moment vector lies in
the interior of the moment space: the base Chebyshev system represents mass, $-U$, and $V$
on the compact interval $[A,\lambda_1]$, and the contact equations make the candidate and
input share these three moments. Lemma~2.1 of Dette--Schorning selects the lower principal
representation containing the left endpoint $A$ and one interior node $c$ as the unique
extremal representation for the appended function $-\operatorname{erfc}(y\sqrt{\lambda})$
\cite{detteschorning2013principal}. The lower principal representation minimizes the appended
negative-tail functional among representations with the matched base moments, so
$-T_g(y)\leq -T_f(y)$ and hence $T_g(y)\geq T_f(y)$. This principal
representation is fixed by the base
moments and hence does not depend on the comparison value $y$, so the resulting candidate
satisfies
\begin{equation}
 T_g(y)\geq T_f(y)\quad\forall y\geq0,
 \qquad\text{i.e.,}\qquad f\preceq g.
 \label{eq:global-safety}
\end{equation}
For $y>0$, $y\neq x_0$, the strict extended-Chebyshev sign makes the input's distinct-node
representation strictly larger for the negative-tail functional, yielding $D(y)>0$. Applying
the same lower-principal mapping to the separate curvature functional gives
$D''(x_0)>0$. The Gaussian determinant checks and these strictness statements are collected
in Appendix~\ref{app:derivations}.

\subsection{Reduction operator}

The single-step operation used later is
\begin{equation}
 C_\alpha[f]=
 \begin{cases}
 f,&K\leq2,\\
 g_{a,p},&K\geq3,
 \end{cases}
 \label{eq:c-operator}
\end{equation}
where the nontrivial branch solves~\eqref{eq:pl-definition}, then~\eqref{eq:r-R}, and
finally~\eqref{eq:parameter-recovery}. It returns at most two components. In exact arithmetic,
\begin{equation}
 f\preceq C_\alpha[f],\qquad
 \operatorname{PL}_\alpha(C_\alpha[f])=\operatorname{PL}_\alpha(f).
 \label{eq:c-properties}
\end{equation}
Thus $C_\alpha$ returns a state with at most two Gaussian components, preserves the current
protection-level contact, and preserves the all-threshold tail order.
Production evaluation uses stable erfc tails, bracketed scalar roots, and a numerical guard on
the recovered weight. The guard moves a finite-precision result toward the conservative side;
Section~V reports raw and guarded diagnostics separately, while the analytic safety statements
are made in exact arithmetic.

%% file: sections/04-sequential.tex
\section{Safe Mixture Reduction for Linear Error Propagation}
\label{sec:sequential}

\subsection{Sequential two-component reduction}

Let $F_i$ denote the exact distribution after the first $i$ inputs in~\eqref{eq:linear-error}
have been processed, and let $G_i$ denote the state retained by the sequential algorithm. The
first nonzero coefficient initializes the states as
\begin{equation}
 F_i=H_i,\qquad G_i=C_\alpha[H_i].
 \label{eq:seq-init}
\end{equation}
Zero coefficients before this initialization are skipped. After initialization, a zero
coefficient is an identity step, $F_i=F_{i-1}$ and $G_i=G_{i-1}$. At a later nonzero input,
\begin{equation}
 F_i=F_{i-1}*H_i,\qquad G_i=C_\alpha[G_{i-1}*H_i].
 \label{eq:seq-recursion}
\end{equation}
If the normalized convolution has at most two distinct scales, $C_\alpha$ is the identity and
the convolution result is retained directly. Thus the state always represents the same prefix
as the exact path, while the stored representation has at most two components.

Figure~\ref{fig:sequential-algorithm} collects the executed update in the order used by the
experiments: step 1 handles zero coefficients, while steps 2 through 5 handle each active update.
The exact prefix
$F_i$ remains an analytical reference outside this procedure.
\begin{figure}[t]
\centering
\fbox{\begin{minipage}{\dimexpr\columnwidth-2\fboxsep-2\fboxrule\relax}
\footnotesize
\textbf{Input:} ordered coefficients $s_i$, component mixtures $H_i$, and selected tail level $\alpha$.\\
\textbf{Output:} retained state $G_N$ with at most two components.
\begin{enumerate}
\item Process $i=1,\ldots,N$ in the original order. If $s_i=0$ and the state is uninitialized, skip to the next input. If $s_i=0$ after initialization, set $G_i=G_{i-1}$ and skip to the next input.
\item For a nonzero coefficient, enter the active-update steps two through five: set $f=H_i$ for the first active input and $f=G_{i-1}*H_i$ thereafter. Normalize $f$ and merge equal scales.
\item If $f$ has $K\leq2$ distinct input scales, set $G_i=f$ and continue to the next input.
\item Otherwise set $b$ to the largest scale in $f$, solve the current-input PL equation~\eqref{eq:pl-definition}, solve the internal-scale root~\eqref{eq:r-R}, and recover the raw $a$ and $p$ from~\eqref{eq:parameter-recovery}.
\item In exact arithmetic, form $G_i=C_\alpha[f]$ with the raw parameters. In floating point, first apply the stated guard to $p$ in this nontrivial branch; if the guarded value leaves $(0,1)$, report reduction failure and return no state, otherwise form $G_i$ with the guarded weight.
\end{enumerate}
\end{minipage}}
\caption{Sequential two-component reduction.}
\label{fig:sequential-algorithm}
\end{figure}

There are three useful propagation paths. Exact keeps
\begin{equation}
 F_N=H_1*\cdots*H_N,
 \label{eq:exact-path}
\end{equation}
terminal C first forms this complete mixture and then applies $C_\alpha$, whereas sequential
C applies~\eqref{eq:seq-recursion} after each nonzero convolution. In exact arithmetic, the
single terminal reduction has the contact identity
\begin{equation}
 \mathrm{P\!L}_\alpha(C_\alpha[F_N])=\mathrm{P\!L}_\alpha(F_N).
 \label{eq:terminal-contact}
\end{equation}
If the next input has $K_i$ effective modes, the sequential convolution has at most $2K_i$
nominal components before reduction. In contrast, the explicit exact path grows with the
product of the effective input mode counts; repeated variances may reduce the distinct-scale
count but do not remove the combinatorial mechanism. Sorting, merging equal scales, evaluating
the input protection level, and solving the internal-scale equation contribute per-step costs whose iteration
counts depend on the numerical brackets, tolerance, and scale separation. The detailed
conditioned accounting is given in Appendix~\ref{app:accounting}.

\subsection{Preservation of tail bounds}

The single-step result~\eqref{eq:c-properties} is combined with the classical convolution
order~\eqref{eq:convolution-order}. Suppose the induction hypothesis is
$F_{i-1}\preceq G_{i-1}$. Since $H_i$ is an independent member of the centered finite
positive-weight Gaussian-mixture class, its density is symmetric and strictly unimodal, and
\begin{equation}
 F_i=F_{i-1}*H_i\preceq G_{i-1}*H_i.
 \label{eq:seq-conv-order}
\end{equation}
The convolved candidate remains a finite centered positive-weight Gaussian mixture, so Section~III
gives
\begin{equation}
 G_{i-1}*H_i\preceq C_\alpha[G_{i-1}*H_i]=G_i.
 \label{eq:seq-c-order}
\end{equation}
Combining~\eqref{eq:seq-conv-order} and~\eqref{eq:seq-c-order} proves
\begin{equation}
 F_i\preceq G_i\quad\text{for every processed prefix},
 \qquad F_N\preceq G_N,
 \label{eq:seq-safety}
\end{equation}
and~\eqref{eq:order-pl} supplies the final protection-level upper bound. The initialization and identity steps
handle the zero-coefficient boundary without applying the density theorem to a zero-scale atom.

\subsection{Conservatism in sequential reduction}

The local contact identity does not imply a globally exact sequential chain. Consider one
nontrivial reduction and call its input and output $f$ and $g$, respectively. Define
\begin{equation}
 D(u)=T_g(u)-T_f(u)\geq0.
 \label{eq:tail-gap}
\end{equation}
For a later independent, nondegenerate kernel $H$ from the same centered finite
positive-weight Gaussian-mixture class, let $p_H$ denote its symmetric density. Integration
by parts gives
\begin{equation}
 T_{g*H}(x)-T_{f*H}(x)
 =\int_0^\infty\bigl[p_H(|x-u|)-p_H(x+u)\bigr]D(u)\,\mathrm{d}u.
 \label{eq:gap-propagation}
\end{equation}
The bracket is the change, per unit shift of the preceding error magnitude $u$, in the
conditional exceedance probability at the fixed threshold $x$. Because $|x-u|<x+u$ and a
Gaussian scale mixture is strictly decreasing on the positive half-line, the bracket is
positive for $u>0$. Hence a tail gap that is positive on a set of positive measure becomes
positive for every $x>0$ after this later convolution.

At $x=\mathrm{P\!L}_\alpha(f*H)$, the reference tail equals $\alpha$, while
\begin{equation}
 T_{g*H}\!\left(\mathrm{P\!L}_\alpha(f*H)\right)>\alpha.
 \label{eq:strict-tail-at-reference}
\end{equation}
Since the candidate tail is strictly decreasing, its threshold must move to the right to return
to $\alpha$, so
\begin{equation}
 \mathrm{P\!L}_\alpha(f*H)<\mathrm{P\!L}_\alpha(g*H).
 \label{eq:strict-pl-gap}
\end{equation}
The nontrivial reduction supplied by Section~III creates the nonnegative gap $D$. The exact
prefix is bounded by the reduced prefix at the next step, and subsequent convolutions preserve
the order; later reductions continue to overbound their current inputs. Therefore a nontrivial
reduction followed by at least one independent nondegenerate convolution is a sufficient
condition for a strict terminal protection-level difference. A reduction at the last step, or an identity
step, does not create this propagation mechanism.

At the current input protection level $x_0$, C has zero tail excess by construction, whereas
the peel loss in~\eqref{eq:peel-loss} is positive for $K\geq3$. This is a comparison at the
selected target point; it does not impose an all-threshold ordering between the two reduced
mixtures. The final 67-to-1 comparison reported in Section~V is therefore an empirical result
over the tested cases.

\subsection{Bounds on Protection Level Inflation}

For an input with $K\geq3$ ordered scales, the two-component sequential peel keeps the widest
component and moves all other mass to the next-widest scale:
\begin{equation}
 g_{\mathrm{peel}}=(1-w_K)\mathcal{N}(0,\sigma_{K-1}^2)+w_K\mathcal{N}(0,\sigma_K^2).
 \label{eq:peel}
\end{equation}
For $K\leq2$, the peel step is the identity and no mass transfer is performed.
Its explicit tail loss is
\begin{equation}
 \begin{aligned}
 \Delta_2(x)&=T_{g_{\mathrm{peel}}}(x)-T_f(x)\\
 &=\sum_{k<K}w_k\left[Q_{\sigma_{K-1}}(x)-Q_{\sigma_k}(x)\right]\geq0.
 \end{aligned}
 \label{eq:peel-loss}
\end{equation}
The general $m$-component expression, for $1\leq m\leq K$, is deferred to Appendix~\ref{app:derivations}; the two-component form is the
main-text baseline for a same-budget, same-timing comparison.

For any normalized centered finite Gaussian-mixture pair $f\preceq g$, define
\begin{equation}
 x_f=\mathrm{P\!L}_\alpha(f),\qquad x_g=\mathrm{P\!L}_\alpha(g),\qquad
 \delta=T_g(x_f)-\alpha.
 \label{eq:pl-certificate-def}
\end{equation}
Because $T_g(x_g)=\alpha$ and $T_g'$ is negative,
\begin{equation}
 \delta=\int_{x_f}^{x_g}|T_g'(t)|\,\mathrm{d}t.
 \label{eq:delta-integral}
\end{equation}
The derivative magnitude is positive and nonincreasing on the positive half-line. Therefore
\begin{equation}
 |T_g'(x_g)|(x_g-x_f)\leq\delta\leq|T_g'(x_f)|(x_g-x_f),
 \label{eq:certificate-sandwich}
\end{equation}
and
\begin{equation}
 \frac{\delta}{|T_g'(x_f)|}\leq x_g-x_f\leq
 \frac{\delta}{|T_g'(x_g)|}.
 \label{eq:pl-certificate}
\end{equation}
This is an offline interpretation tool: it requires both protection-level values, the candidate tail at the
reference protection level, and the candidate endpoint derivatives. It can be applied to $(F_N,G_N)$, a
local pair $(G_{i-1}*H_i,G_i)$, or $(f,g_{\mathrm{peel}})$. For an exact local C step,
$\delta=0$ in exact arithmetic; finite-precision guards are reported separately in Section~V.

%% file: sections/05-results.tex
\section{Numerical Evaluation}
\label{sec:results}

The evaluation follows the propagation question posed in Section~II. The main matrix contains four input families, $N\in\{2,4,8,12\}$, and $\alpha\in\{10^{-5},10^{-7},10^{-9}\}$, giving 48 cases; 12 held-out cases and eight boundary cases extend the same implementation checks to 68 cases in total. A separate 16-case matrix spans two families, two geometries, $N\in\{8,12\}$, and $\alpha\in\{10^{-7},10^{-9}\}$. The four representative supplement rows in this section use $N=12$, $\alpha=10^{-7}$. All timings use the recorded Python~3.11 environment, include model construction, propagation, and the final protection-level calculation, and average three repetitions; independent erfc and raw-tail diagnostics are evaluated outside the timed region.

The four main families are baseline, heavy-tail, wide-tail, and heterogeneous mixtures. Their
baseline, heavy-tail, and wide-tail component $(w,\sigma)$ pairs are respectively
$(0.99,1),(0.00999999,3),(10^{-8},12)$; $(0.99,1),(0.0099,3),(10^{-4},12)$; and
$(0.99,1),(0.00999999,3),(10^{-8},50)$. Heterogeneous inputs cycle through the four
templates listed in Appendix~\ref{app:accounting}. The projection coefficients start at one,
alternate sign, and increase in magnitude by successive factors of $\sqrt{1.07}$. The held-out
and boundary configurations include single- and double-scale inputs, repeated scales, zero
weights or projections, nearly coincident scales, small tail weights, and small $\alpha$.
The supplementary matrix uses a designed diagonal precision matrix and the unnormalized
vertical projection rows of the complete synthetic weighted least-squares (WLS) geometries;
its strong and weak labels describe two levels of geometric conditioning pressure. The recorded
condition numbers of $G^{\mathsf T}WG$ are approximately $6\times10^1$--$7\times10^1$ for
strong geometries and $6.8\times10^6$--$3.6\times10^7$ for weak geometries.
The strong sky uses dispersed azimuth/elevation rows, whereas the weak sky uses near-collinear
rows; the complete $G$, $W$, and projection matrices are included with the ancillary records.

The analytic single-Gaussian reference (SG) uses the propagated maximum-scale variance and the
same two-sided Gaussian tail equation.

\begin{table}[t]
\caption{Numerical checks over the 68-case main system (272 model evaluations). Component counts are the largest observed nominal counts before and after sequential reduction. The three residual rows refer to distinct diagnostics.}
\label{tab:checks}
\centering
\begin{tabular}{l l}
\toprule
Quantity & Recorded value \\
\midrule
Cases / model evaluations & 68 / 272 \\
Maximum sequential pre-reduction components & 6 \\
Maximum sequential post-reduction components & 2 \\
Maximum absolute raw local PL residual [m] & $6.253\times10^{-13}$ \\
Maximum absolute independent-erfc PL residual [m] & $6.111\times10^{-13}$ \\
Maximum absolute terminal raw residual [m] & $2.700\times10^{-13}$ \\
Count with $\mathrm{PL}_{\rm candidate}<\mathrm{PL}_{\rm exact}-10^{-8}\,\mathrm{m}$ & 0 \\
\bottomrule
\end{tabular}
\end{table}

Table~\ref{tab:checks} separates three signed diagnostics; the displayed residual values are
their maximum absolute differences. The local raw residual is
$\mathrm{PL}_{\rm local,before\ guard}-x_0$ at each sequential C step; the independent-erfc
residual is $\mathrm{PL}_{\rm erfc\ bisection}-\mathrm{PL}_{\rm reported}$ for each saved
model check; and the terminal raw residual is
$\mathrm{PL}_{\rm terminal,before\ guard}-\mathrm{PL}_{\rm exact}$ on the main cases.
Across the 68 cases and 272 method evaluations, the acceptance rule counts a candidate
underbound when $\mathrm{PL}_{\rm candidate}<\mathrm{PL}_{\rm exact}-10^{-8}\,\mathrm{m}$;
the count is zero. The independent-erfc residual maximum is taken over the 256 saved checks
(204 candidate models and 52 exact models, with exact checks limited to the smaller recorded
mixtures); the terminal and local residuals use their stated populations. The analytic safety
statement is the induction in Section~IV.

\begin{table*}[t]
\caption{Representative protection levels in metres for $N=12$ and $\alpha=10^{-7}$. The analytic single-Gaussian column uses the propagated maximum component scale and the same two-sided tail equation. Exact and terminal C agree to the displayed precision in the reported cases.}
\label{tab:pl-main}
\centering
\begin{tabular}{lrrrrr}
\toprule
Case & Exact PL [m] & Sequential C PL [m] & Sequential peel PL [m] & Terminal C PL [m] & Analytic SG PL [m] \\
\midrule
Baseline, $N=12$, $\alpha=10^{-7}$ & 29.125 & 36.974 & 263.171 & 29.125 & 270.351 \\
Wide-tail maximum ratio & 30.861 & 68.705 & 1094.640 & 30.861 & 1126.462 \\
\bottomrule
\end{tabular}
\end{table*}

For the baseline case, the sequential C value is 36.974~m, whereas the fixed peel is 263.171~m; the terminal C value returns to 29.125~m because its reduction is applied only after the complete exact convolution. The analytic single-Gaussian value is 270.351~m for the same propagated maximum-scale reference. Across the 48 main cases, the sequential-C-to-exact PL ratio ranges from 1.00076 to 2.22627. The largest ratio occurs in the wide-tail $N=12$, $\alpha=10^{-7}$ case shown in Table~\ref{tab:pl-main}. Over all 68 cases, sequential C is tighter than sequential peel in 67 cases and equal in one boundary case. This pattern is consistent with the mechanism in Section~IV: peel transfers all non-widest mass to the next-widest scale, while C re-solves the internal scale and wide-component weight at the current input protection level.

\begin{figure*}[t]
\centering
\includegraphics[width=0.94\textwidth]{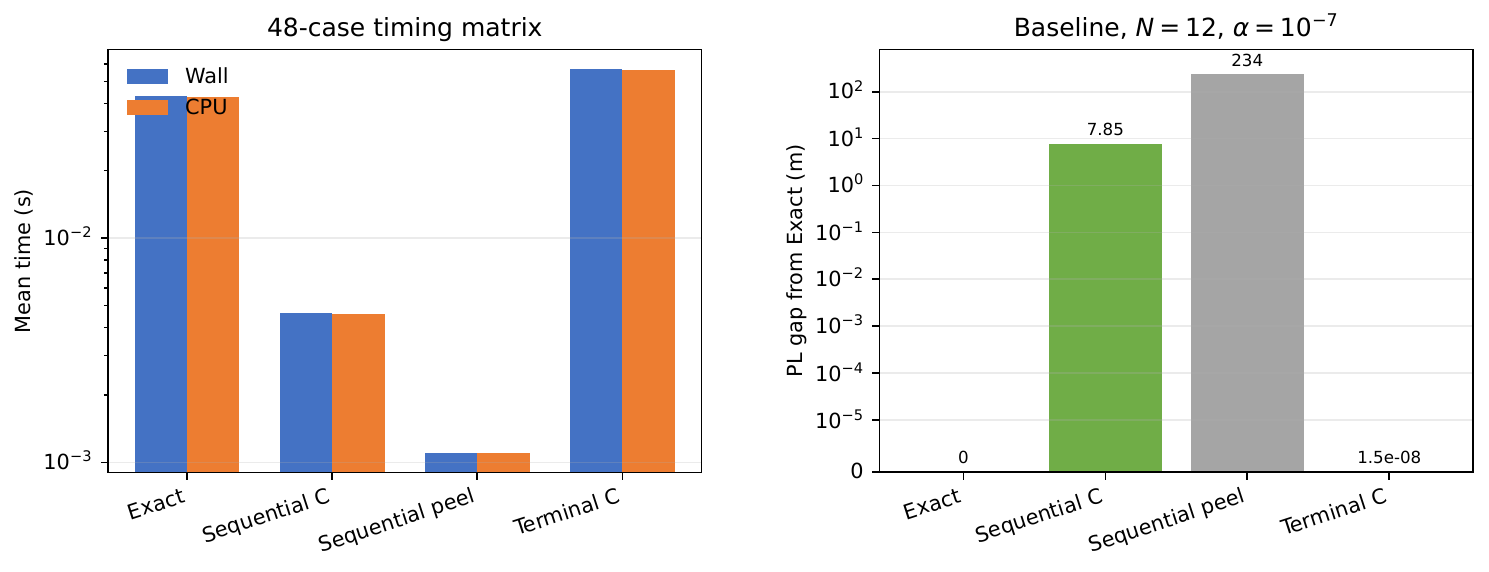}
\caption{(a) Mean wall and CPU times over the 48 main cases, with one value for each propagation method. (b) Protection-level gaps for the baseline $N=12$, $\alpha=10^{-7}$ case.}
\label{fig:main-summary}
\end{figure*}

Figure~\ref{fig:main-summary}(a) summarizes the timing and Figure~\ref{fig:main-summary}(b) shows the corresponding representative PL gaps. The mean wall times for exact, sequential C, sequential peel, and terminal C are 42.95, 4.66, 1.10, and 57.17~ms, respectively, over the 48-case timing matrix. The representative gaps are 7.849~m for sequential C, 234.046~m for sequential peel, and $1.50\times10^{-8}$~m for terminal C after guarding. The timing and PL panels therefore describe different aspects of the same propagation choices: the compact state reduces the retained mixture size, while a reduction performed before later convolutions carries the tail gap analyzed in Section~IV.

For the same-budget comparison in Table~\ref{tab:supplement}, the comparator is a
two-component upward merge. Sort the input scales as $\sigma_1\leq\cdots\leq\sigma_K$ and enumerate
the contiguous cuts $j=1,\ldots,K-1$; for $K\leq2$, use the identity. For each cut
with $K\geq3$, replace the left group by one centered Gaussian component at
$\sigma_j$ carrying its aggregated weight, and replace the right group by one
centered Gaussian component at $\sigma_K$ carrying its aggregated weight. Both
replacement scales are therefore no smaller than the scales they represent. Select
the candidate with the smallest output PL at the current reducer input, resolving
ties by the smallest cut index $j$. This is the fixed benchmark selector used here
to instantiate Gao et al.'s variance-upward principle~\cite{gao2022gmm}, not an algorithm prescribed
by that paper.

\begin{table*}[t]
\caption{Four representative rows from the 16-case same-budget supplement at $N=12$ and $\alpha=10^{-7}$. The upward-merge method follows the variance-upward principle described by Gao et al.; the experiment instantiates that principle with a fixed two-component contiguous split and PL selection.}
\label{tab:supplement}
\centering
\begin{tabular}{llrrrr}
\toprule
Family & Geometry & Exact PL [m] & Sequential C PL [m] & Upward-merge PL [m] & Analytic SG PL [m] \\
\midrule
Ordinary & Strong & 12.540 & 15.084 & 25.153 & 99.789 \\
Ordinary & Weak & 1680.848 & 1980.696 & 3189.815 & 12655.015 \\
Wide-tail & Strong & 13.047 & 18.738 & 26.687 & 415.789 \\
Wide-tail & Weak & 1730.868 & 2089.272 & 3384.386 & 52729.227 \\
\bottomrule
\end{tabular}
\end{table*}

\begin{table*}[!b]
\caption{Mean wall time and explicit mixture size by $N$ in the 48-case main timing matrix. Each row averages 12 cases and three repetitions per method. Times are in milliseconds.}
\label{tab:timing}
\centering
\begin{tabular}{rrrrrrrr}
\toprule
$N$ & Exact wall [ms] & Sequential C wall [ms] & Peel wall [ms] & Terminal C wall [ms] & Exact terms [count] & C pre [count] & C post [count] \\
\midrule
2 & 0.524 & 1.850 & 0.748 & 1.127 & 9 & 6 & 2 \\
4 & 0.593 & 2.828 & 0.858 & 1.193 & 81 & 6 & 2 \\
8 & 3.221 & 5.315 & 1.159 & 4.330 & 6561 & 6 & 2 \\
12 & 167.469 & 8.666 & 1.651 & 222.013 & 531441 & 6 & 2 \\
\bottomrule
\end{tabular}
\end{table*}

Table~\ref{tab:supplement} gives four representative rows from the separate supplement. In all 16 supplementary records, sequential C has a lower final PL than this fixed two-component comparator. The difference follows from the parameterization: upward merging keeps the selected input scales and aggregates group weights, whereas C retains the current largest scale and solves for the internal scale from the current PL before recovering the wide weight. The analytic single-Gaussian values use the same maximum-scale variance formula and are shown as a reference representation, not as an additional timing method.

Table~\ref{tab:timing} resolves the scale dependence of the runtime comparison. For $N=2,4,8$, the C construction is slower than exact evaluation in this environment because the explicit mixture remains small. At $N=12$, exact propagation retains 531441 nominal terms and averages 167.469~ms, while sequential C averages 8.666~ms with at most six components before each reduction and two afterward. Terminal C averages 222.013~ms because it retains the complete mixture until the final step. The separate 16-case supplement has a mean sequential-C time of approximately 21.65~ms and a fixed upward-merge time of approximately 79.75~ms; those timings use its own input matrix and are not combined with the main timing rows.

The analytic single-Gaussian (SG) reference is defined by the maximum-scale variance after propagation, $\sigma_{\mathrm{SG}}^2=\sum_i s_i^2\sigma_{i,\max}^2$, followed by the same two-sided Gaussian tail equation. This deterministic reference supplies the analytic SG columns in Tables~\ref{tab:pl-main} and~\ref{tab:supplement}; no matched single-Gaussian timing is reported because that method was not timed in the accepted records.

The complete machine-readable records accompany the source package as ancillary data. Gao et al.'s variance-upward rule is stated in the discussion following printed p.~12, Eq.~(17)~\cite{gao2022gmm}; the contiguous two-group enumeration and final PL selection above are the comparator specified for this evaluation. For the supplementary geometries, $S=(G^{\mathsf T}WG)^{-1}G^{\mathsf T}W$; the unnormalized vertical row of $S$ supplies the projection coefficients, and $W$ is a designed diagonal precision matrix.

%% file: sections/06-conclusion.tex
\section{Conclusion}
\label{sec:conclusion}

This paper develops a compact safety-preserving representation for centered finite Gaussian mixtures propagated through independent scalar linear combinations. Fixing the largest input scale and solving the Gaussian-tail contact equations gives a calculable two-component candidate; the principal-representation result supplies its all-threshold tail order, and convolution preservation carries that order through sequential propagation. The contact is local to the current input, so a nontrivial reduction followed by a later independent nondegenerate convolution can create a strict terminal PL gap; the tail-gap identity and the endpoint-derivative certificate explain how that gap is formed and measured.

The synthetic results connect this mechanism to computational use. In the separate 16-case same-budget matrix, sequential C has a lower PL than the fixed two-component upward-merge comparator in all 16 records. In the 48-case main timing matrix, the $N=12$ layer averages 8.67~ms for sequential C and 167.47~ms for exact propagation, while at smaller $N$ the explicit mixtures remain small enough that Exact is cheaper in the recorded environment. The two evidence sets therefore complement one another: the supplement measures PL tightness at a common budget, and the main matrix shows when the compact state offsets explicit mode growth. Together they support a compact representation for continued integrity-oriented error propagation in the stated synthetic setting.

Two extensions follow directly from the resulting representation. An application study can insert the sequential state into a complete navigation-integrity chain and examine how observation geometry, model uncertainty, and risk allocation interact with the propagated tail bound. A theoretical study can extend the reduction and propagation conditions to non-centered or asymmetric inputs, correlated input errors, and larger representation budgets. These directions retain the central question established here: how to reduce an explicit error model while preserving a checkable tail relation for the computation that follows.

%% file: sections/07-appendices.tex
\section{Additional Derivations}
\label{app:derivations}

\subsection{Convolution closure and the principal-representation specialization}
\label{app:closure}

For centered symmetric laws, the two-sided tail order in~\eqref{eq:tail-order} is the peakedness order. The classical convolution theorem states that convolution with an independent symmetric-unimodal law preserves this order; the Gaussian-mixture kernels in this paper satisfy the required condition. For completeness, let $r_x(u)=\Pr(|u+Z|>x)$ for $Z\sim H$. Differentiating the two tail boundaries gives
\begin{equation}
r_x'(u)=p_H(|x-u|)-p_H(x+u),
\end{equation}
where the derivative is understood piecewise at $u=x$. Symmetry and unimodality imply $r_x'(u)\geq0$. If $q=p_g-p_f$ and $D(u)=T_g(u)-T_f(u)$, then $D(u)=2\int_u^\infty q(v)\,\mathrm{d}v$ on $u\geq0$, and integration by parts gives~\eqref{eq:gap-propagation}. The boundary terms vanish because $D(0)=D(\infty)=0$. Strict decrease of $p_H$ makes the integrand positive for $u>0$ whenever $H$ is nondegenerate, so a nonzero nonnegative $D$ becomes positive after the later convolution.

The principal-representation step in Section~III uses the Gaussian-tail functions as a Chebyshev system in the precision variable $\lambda$. Define
\begin{equation}
d(\lambda)=-U'(\lambda)=\frac{x_0}{\sqrt{\pi\lambda}}e^{-\lambda x_0^2}.
\end{equation}
For the base system $\{1,-U(\lambda),V(\lambda)\}$, the first three Wronskians reduce to
\begin{equation}
W_1=1,\qquad W_2=d(\lambda),\qquad W_3=2x_0d(\lambda)^2>0.
\end{equation}
The input nodes lie in the compact interval $[A,\lambda_1]$. For $K\geq3$ distinct nodes
with positive weights, the three matched quantities—mass, the $-U$ moment, and the $V$
moment—form an interior point of the corresponding moment space. The contact equations in
Section~III match these moments for the input and the candidate. Lemma~2.1 of
Dette--Schorning then maps this interior moment vector to the lower principal representation
with the left endpoint $A$ and one interior node $c$; the representation is unique for the
appended extremal functional.
For a comparison threshold $y>0$, $y\neq x_0$, the extended determinant can be written as
\begin{equation}
W_4=2x_0d(\lambda)^3\left(\frac{\Omega'(\lambda)}{d(\lambda)}\right)''\!,
\end{equation}
where $\Omega(\lambda)=-\operatorname{erfc}(y\sqrt{\lambda})$. In fact,
\begin{equation}
\begin{aligned}
\frac{\Omega'(\lambda)}{d(\lambda)}&=\frac{y}{x_0}e^{-(y^2-x_0^2)\lambda},\\
\left(\frac{\Omega'(\lambda)}{d(\lambda)}\right)''&=
\frac{y}{x_0}(y^2-x_0^2)^2e^{-(y^2-x_0^2)\lambda}>0.
\end{aligned}
\end{equation}
The strict sign holds on $[A,\lambda_1]$ for $y>0$, $y\neq x_0$; at $y=x_0$ the added functional repeats the base tail constraint, and at $y=0$ normalization handles the endpoint. Because the lower principal representation is determined by the three base moments, the same two-node candidate is used for every comparison value $y$. Lemma~2.1 therefore gives $D(y)>0$ for $y>0$, $y\neq x_0$, when the input has additional distinct positive-weight nodes; the separate curvature functional is handled by the same mapping and yields $D''(x_0)>0$.

The contact curvature can also be checked directly. With $h(\lambda)=U(A)-U(\lambda)$ and $\psi(\lambda)=V(A)-V(\lambda)$, the auxiliary ratio satisfies
\begin{equation}
\frac{\psi'(\lambda)}{h'(\lambda)}=\frac{V'(\lambda)}{U'(\lambda)}=\frac{1}{x_0}-2x_0\lambda,
\end{equation}
For the curvature calculation, set the separate curvature functional to the second derivative
of the Gaussian tail at the contact point:
\begin{equation}
\begin{aligned}
H(\lambda)&=\left.\partial_y^2\operatorname{erfc}(y\sqrt{\lambda})\right|_{y=x_0}\\
&=\frac{4x_0\lambda^{3/2}}{\sqrt{\pi}}e^{-\lambda x_0^2},
\qquad \Omega_c(\lambda)=-H(\lambda).
\end{aligned}
\end{equation}
Then
\begin{equation}
\frac{\Omega_c'(\lambda)}{d(\lambda)}=\frac{H'(\lambda)}{U'(\lambda)}=4x_0^2\lambda^2-6\lambda,
\qquad \left(\frac{\Omega_c'}{d}\right)''=8x_0^2.
\end{equation}
The difference between the candidate and input weighted sums of $H$ is exactly the curvature
difference $D''(x_0)$, so the nondegenerate contact consequently has $D''(x_0)>0$. For the Jensen step, define
\begin{equation}
\bar h=\frac{\sum_{k<K}w_kh(\lambda_k)}{1-w_K},\qquad
\bar\psi=\frac{\sum_{k<K}w_k\psi(\lambda_k)}{1-w_K}.
\end{equation}
Strict concavity of $P$ gives $P(\bar h)>\bar\psi$. The contact equation gives $\bar\psi/\bar h=R=r(c)$, so the decreasing ratio $r=\psi/h$ implies $\bar h<h(c)$ and hence
\begin{equation}
1-p=\frac{(1-w_K)\bar h}{h(c)}<1-w_K.
\end{equation}
Together with $1-p>0$, this places the recovered wide weight in the admissible interval.

For a general $m$-component peel with ordered scales and $1\leq m\leq K$, the retained nodes are the widest $m-1$ scales and all lighter mass is transferred to the next retained scale:
\begin{equation}
\begin{aligned}
g_{\mathrm{peel}}^{(m)}={}&\sum_{k=K-m+2}^{K}w_k\phi(\cdot;0,\sigma_k^2)\\
&+\left(\sum_{k\le K-m+1}w_k\right)\phi(\cdot;0,\sigma_{K-m+1}^2).
\end{aligned}
\end{equation}
Its pointwise tail loss is
\begin{equation}
\Delta_m(x)=\sum_{k\le K-m+1}w_k\left[Q_{\sigma_{K-m+1}}(x)-Q_{\sigma_k}(x)\right]\geq0.
\end{equation}
Writing $W_m=\sum_{k\le K-m+1}w_k$ gives
\begin{equation}
\begin{aligned}
\Delta_m(x)-\Delta_{m+1}(x)
&=W_{m+1}Q_{\sigma_{K-m+1}}(x)\\
&\quad-W_{m+1}Q_{\sigma_{K-m}}(x)\geq0,
\end{aligned}
\end{equation}
so the explicit loss decreases to zero as $m$ reaches $K$; the difference formula applies for $1\leq m<K$.

\section{Implementation and Recorded Data}
\label{app:accounting}

\begin{table*}[t]
\caption{Complete 16-case same-budget supplement. Values are protection levels in metres; SG is the analytic maximum-scale Gaussian reference.}
\label{tab:all-supplement}
\centering\footnotesize
\begin{tabular}{llrrr rrr}
\toprule
Family & Geometry & $N$ & $\alpha$ & Exact PL [m] & Sequential C PL [m] & Upward $m=2$ PL [m] & Analytic SG PL [m] \\
\midrule
ordinary & strong & 8 & $10^{-7}$ & 14.432 & 17.729 & 28.453 & 113.210 \\
ordinary & strong & 8 & $10^{-9}$ & 21.896 & 24.485 & 52.070 & 129.844 \\
ordinary & strong & 12 & $10^{-7}$ & 12.540 & 15.084 & 25.153 & 99.789 \\
ordinary & strong & 12 & $10^{-9}$ & 18.023 & 20.413 & 47.575 & 114.452 \\
ordinary & weak & 8 & $10^{-7}$ & 6266.732 & 6866.520 & 9519.020 & 37874.900 \\
ordinary & weak & 8 & $10^{-9}$ & 9093.154 & 9399.863 & 17057.768 & 43440.078 \\
ordinary & weak & 12 & $10^{-7}$ & 1680.848 & 1980.696 & 3189.815 & 12655.015 \\
ordinary & weak & 12 & $10^{-9}$ & 2521.653 & 2711.379 & 5933.965 & 14514.489 \\
wide-tail & strong & 8 & $10^{-7}$ & 14.816 & 19.017 & 29.160 & 471.708 \\
wide-tail & strong & 8 & $10^{-9}$ & 89.926 & 98.062 & 211.999 & 541.019 \\
wide-tail & strong & 12 & $10^{-7}$ & 13.047 & 18.738 & 26.687 & 415.789 \\
wide-tail & strong & 12 & $10^{-9}$ & 73.748 & 80.537 & 196.864 & 476.884 \\
wide-tail & weak & 8 & $10^{-7}$ & 6352.417 & 7084.006 & 9755.457 & 157812.083 \\
wide-tail & weak & 8 & $10^{-9}$ & 37712.836 & 38444.206 & 66405.758 & 181000.324 \\
wide-tail & weak & 12 & $10^{-7}$ & 1730.868 & 2089.272 & 3384.386 & 52729.227 \\
wide-tail & weak & 12 & $10^{-9}$ & 10381.543 & 10788.971 & 24252.528 & 60477.037 \\
\bottomrule
\end{tabular}
\end{table*}

After the first nonzero input, each sequential state has at most two components. Convolution with an input having $K_i$ effective modes produces at most $2K_i$ nominal components before reduction; sorting and equal-scale aggregation cost $O(m_i\log m_i)$ for $m_i$ pre-reduction terms, while the PL and internal-scale roots contribute iteration-dependent costs. A closed bracket, positive scale separation, and the recorded tolerances are the conditions under which these root counts remain bounded. Near-coincident scales use the 180-digit fallback recorded by the implementation. The exact path has a nominal product $\prod_{i\in I}K_i$ of mode combinations. Numerically, the recovered raw weight is shifted to $p=p_{\rm raw}+8\,\varepsilon_{\rm mach}$ toward the conservative side; an output with $p\geq1$ is reported as a failure rather than clamped. This guard is not a directed-rounding certificate. The recorded timing includes propagation and final protection-level evaluation, whereas independent erfc and raw-tail diagnostics are outside the timed region.

The main timing matrix contains 48 cases, four methods, and three repetitions, giving 576 raw timing rows. The 12 held-out and eight boundary cases contribute model checks rather than rows in that timing matrix. The supplement contains 16 records with the same two-component budget for sequential C and the fixed upward-merge comparator; its timing has three repetitions and 48 raw rows per method. Its complete protection-level rows are listed in Table~\ref{tab:all-supplement}.

The heterogeneous family cycles through the following four centered input templates in order:

\begin{table}[t]
\caption{Heterogeneous input templates used cyclically in the main matrix.}
\label{tab:heterogeneous-templates}
\centering
\footnotesize
\begin{tabular}{c l}
\toprule
Template & $(w,\sigma)$ pairs \\
\midrule
1 & $(0.965,0.65),(0.034,2.1),(0.001,6.5)$ \\
2 & $(0.91,0.9),(0.088,2.8),(0.002,8.0)$ \\
3 & $(0.985,1.2),(0.014,3.7),(0.001,10.0)$ \\
4 & $(0.94,0.75),(0.058,2.4),(0.002,5.5)$ \\
\bottomrule
\end{tabular}
\end{table}

The standard run used Python~3.11.15, NumPy~2.4.2, SciPy~1.17.1, and mpmath~1.3.0 on
macOS~15.5 arm64 with one configured thread. For the supplementary geometries,
$S=(G^{\mathsf T}WG)^{-1}G^{\mathsf T}W$; the unnormalized vertical row of $S$ supplies the
projection coefficients. The matrix $W$ is a designed diagonal precision matrix, and the
complete $G$, $W$, and $S$ records accompany the ancillary data.